\documentclass[submission,copyright,creativecommons]{eptcs}
\providecommand{\event}{WS-FMDS 2012} % Name of the event you are submitting to
\usepackage{times}
\usepackage{enumerate}
\usepackage{epsfig}
\usepackage{ltcadiz}
\usepackage{lscape}

\title{Predictive Software Measures based on Z Specifications -- \\ A Case Study}
\author{Andreas Bollin
\institute{Software Engineering and Soft Computing\\
University of Klagenfurt\\
Klagenfurt, Austria}
\email{Andreas.Bollin@aau.at}
\and
Abdollah Tabareh
\institute{Department of Computer Science and Engineering\\
University of Gothenburg\\
Gothenburg, Sweden}
\email{tabareh@gmail.com}
}
\def\titlerunning{Predictive Software Measures based on Z Specifications -- A Case Study}
\def\authorrunning{A. Bollin, A. Tabareh}

\begin{document}

\maketitle

\begin{abstract}
Estimating the effort and quality of a system is a critical step at the beginning of every software project. It is necessary to have reliable ways of calculating these measures, and, it is even better when the calculation can be done as early as possible in the development life-cycle.

Having this in mind, metrics for formal specifications are examined with a view to correlations to complexity and quality-based code measures. A case study, based on a Z specification and its implementation in $ADA$, analyzes the practicability of these metrics as predictors.
\end{abstract}

%Keywords: Formal Specification, Comprehension, Slices and Chunks

\newtheorem{Def}{\bf Definition}

\section{Introduction}

Recent studies in the areas of software metrics and project management have stimulated a lot of ideas of how development effort can be estimated and which metrics are of relevance \cite{Jorgensen04,putnam03,sneed10}. Basically, they all suggest that the collection of data and the estimation process should be performed as early and as objectively as possible -- so why not taking a closer look at properties of formal specifications?

\medskip

To the best of our knowledge, the only publicly available case study that took a closer look at correlations between specifications and implementations was conducted by Samson, Nevill and Dugard in 1987 \cite{sams87}. The authors used Modula-2 modules and a HOPE specification to show that there is a correlation between the number of equations in HOPE and the number of lines of source code and cyclomatic complexity in the Modula-2 modules. However, the authors admit that the study is relatively small-scale as their data is based on only 9 experimental subjects.

\medskip

The objective of this paper is now to shed some more light onto the question whether specifications' properties can help predicting attributes of derived implementations or not.
For this, the following strategy is pursued: firstly, based on a set of well-known measures, it tries to find out whether some of the measures are correlated or not. Secondly, it suggests a prediction model for some of the measures. A case study, based on the specification and implementation of the $Tokeneer$ system \cite{chap09} forms the basis for these considerations. It takes the Z specification of the system and its implementation in $ADA$ as the point of departure and identifies those parts of the code that unambiguously implement specific parts of the specification. After that, it calculates size, structure and quality related measures for both of the documents. Finally, it looks for correlations between the measures, and, based on the findings, it calculates a prediction model for several $ADA$-based size- and complexity-related  measures.

\medskip

This paper is structured as follows: Section \ref{sec:Measures} briefly introduces the code and specification measures that are used in the study. Section \ref{sec:Study} presents the setting of the study, the experimental subject and the statistical tests used. Next, Section \ref{sec:Correlation} presents and discusses the results of the correlation tests, and Section \ref{sec:Prediction} presents the prediction model. Section \ref{sec:Validity} discusses possible threats to validity, and, finally, Section \ref{sec:Conclusion} summarizes the findings and discusses possible steps to be done next.

\section{Measures}\label{sec:Measures}

This section introduces the measures used for assessing the $Z$ specification and its implementation in $ADA$. Please note that, due to limitations of space, only a brief overview of the measures is provided\footnote{An in-depth discussion of other specification-based measures can be found in the Ph.D. thesis of Bollin \cite{boll04}.}.

\subsection{Code-based Measures}

The implementation language of the $Tokeneer$ specification \cite{chap09} is $ADA$. In his master thesis, Tabareh \cite{tab11} took a look at currently available environments that are able to generate practical measures from $ADA$ code. He suggests to apply the $Understand$ tool and uses the following measures (where $M$ denotes either an $ADA$ function or a procedure)\footnote{The tool and a description of the measures can be found at the Understand homepage at www.scitools.com. Page last visited: May 2012.} for a preliminary study comparing ADA and Z-based measures:

\vspace{-0.2em}
\begin{itemize}\setlength{\itemsep}{-2pt}
\item CountLine $CL(M)$. It counts the number of physical lines.
\item CountLineCode $CLC(M)$. It counts the number of lines that contain source code.
\item CountLineExecutable $CLCE(M)$. It counts the number of lines containing executable $ADA$ code.
\item CountLineCodeDecl $CLCD(M)$. It count the number of lines containing declarative $ADA$ code.
\item Knots Count $KNOTS(M)$: It is a measure for the structuredness of a module and counts overlapping jumps in the program flow graph.
\item Cyclomatic Complexity $CYC(M)$. It measures the maximum number of linearly independent paths through a program and is extracted by counting the minimum set of paths which can be used to construct all other paths through the graph.
\end{itemize}

In order to focus even more on structural properties of the code, this study additionally makes use of Shepperd and Ince's Information flow count \cite{shep90}. The general idea is that the complexity of a module is related to the number of flows or channels of information between the module and its environment. For this, the $Understand$ tool can be used to generate the call-graph, and the flow of data and control then forms the basis for the calculation of the Sheppard Information Flow $SI$ of a module $M$:

\vspace{-0.2em}
\begin{itemize}\setlength{\itemsep}{-2pt}
\item Fan-in ($FIN(M)$): It comprises the number of data-flows terminating at a component $M$.
\item Fan-out ($FOUT(M)$): It comprises the number of data-flows originating at a component $M$.
\item Information Flow($SI(M)$): It comprises the number of information flows related to a component $M$ and is calculated via $(FIN(M)*FOUT(M))^2$.
\end{itemize}

\subsection{Specification Measures}

Most of the complexity measures for formal specifications focus on size. The reasons are that size-based measures (like lines of specification text) are easy to calculate and yield a single number that is easy to interpret. This is not so much the case for structure- and quality-related measures. Their calculation is usually based on the notion of control and data dependencies, concepts that are not necessarily dominant principles of a specification language. However, several authors \cite{chang94,mitt03,oda93} demonstrated that a reconstruction of these dependencies is possible.

Recently, Bollin showed that coupling and cohesion based measures can reasonably be mapped to formal Z specifications \cite{Bollin10}. The basis for the calculation of all the measures is a graph that contains vertices (called $prime$s) for all predicates and declarations of the specification, and arcs representing (reconstructed) control and data dependencies \cite{boll07a}. With such a graph as a basis, the following measures (defined for schemas $\psi$ that are part of a specifications $\Psi$) are used in the remainder of this work:

\vspace{-0.2em}
\begin{itemize}\setlength{\itemsep}{-2pt}
\item Conceptual Complexity $CC(\psi)$: The conceptual complexity equals the total number of prime vertices in the graph (representing a schema $\psi$).
\item Logical Complexity $v'(\psi)=(l,u)$: The lower bound value $l$ of the measure is 1 plus the number of primes that are terminal vertices of control dependency arcs. It can be compared to counting the number of decision statements in programs. The upper bound value $u$ equals 1 plus the total number of control dependencies. It reflects the total amount of dependencies to be considered.
\item Definition-Use Count: $DU(\psi)$: This measure equals the total number of data dependencies.
\item Use Count $USE(\psi)$: The use count equals the number of identifiers used in the schema $\psi$.
\item Definition Count $DEF(\psi)$: The definition count equals the number of identifiers referring to an after state in the schema $\psi$ of the specification.
\item And-Count $AND(\psi)$: This measure equals the number of AND-combined predicates in $\psi$.
\item Or-Count $OR(\psi)$: This measure equals the number of OR-combined predicates in $\psi$.
\end{itemize}

Semantics-based measures can be calculated by generating slices. The idea goes back to the work of Weiser \cite{weiser82} who introduced five slice-based measures for cohesion: Tightness, Coverage, Overlap, Parallelism and Clustering. Ott and Thuss \cite{ott89} partly formalized these measures. Coupling, on the other hand, was originally defined as the number of local information flow entering (fan-in) and leaving (fan-out) a procedure \cite{hen81}. Harman et.~al \cite{har97} demonstrate that it can also be calculated via slicing. Mapping and evaluating their approaches to Z leads to the following set of quality-based specification metrics \cite{Bollin10}:

\vspace{-0.2em}
\begin{itemize}\setlength{\itemsep}{-2pt}
\item $Coverage$ $Cov(\psi)$: It measures the compactness of a schema by comparing the length of all possible slices to the length of the specification schema $\psi$.
\item $Overlap$ $O(\psi)$: It measures the conciseness of a schema $\psi$ by counting those statements that are common to all of the possible slices and relates the number to the size of all slices.
\item $Schema$ $Coupling$ $\chi(\Psi, \psi_i)$: It is the weighted measure of the information flow between a given schema $\psi_i$ and all other schemas in $\Psi$.
\end{itemize}

\section{The Study}\label{sec:Study}

The study is split into two parts and it aims at answering the following two questions: (a) what type of correlations exists between specification-based and code-based measures, and (b) is it possible to predict code-based measures from specification-based measures?

\medskip

The Tokeneer system  \cite{chap09} is one of the rare, industrial-size and publicly available, formal Z specifications that comes along with a fully derived implementation. It has been developed by Praxis and the $NSA$ and provides a specification for an identification station consisting of a fingerprint reader, a display and a card reader.
%The whole Tokeneer system consists of a secure enclave and a set of system components, some housed inside the enclave and some outside.
The code, written in $ADA$, consists of $11,807$ lines of executable $ADA$ code ($34,769$ lines including comments). The exceptional feature of the $ADA$ files is that they contain so-called ``trace unit'' comments which are direct links to the corresponding sections in the formal design document, thus linking specification text (schemas) and implementation code pairs (procedures and functions) unambiguously together.
%This document makes use of Z in order to specify the formal design units. Figure \ref{fig:Tokeneer} demonstrates this link between the formal specification and the implementation. In this example the $ADA$ specification file contains a ``traceto'' comment referring to the $FD.Admin.AdminStartOp$ section in the design document. It also contains a ``traceunit'' comment referring to the corresponding $ADA$ database file ($admin.adb$) and the procedure called $StartOp$.
The Z specification consists of $11,356$ lines of text, including $4,808$ lines of specification text. The specification itself contains $3,295$ declarations and predicates, it contains $132,088$ control dependencies and $6,145$ data-dependencies.

\medskip

%\begin{figure}[t]
%  \centerline{\psfig{figure=Tokeneer.eps,width=0.9\textwidth}}
%  \caption{Snapshot of the Functional Design document (to the left) with unique identifiers for the %units and the corresponding entry in the $ADA$ specification file (to the right).
%    \label{fig:Tokeneer}}
%\end{figure}

The subjects for this study are set of pairs of code modules (procedures and functions) and their related Z specification (schemas)\footnote{The set of experimental subjects can be found in an Appendix (containing relevant background materials) at the $ViZ$ \cite{boll07a} homepage via the link http://viz.uni-klu.ac.at/images/research/materials/fmds12-addon.pdf. Page last visited: May 2012.}. However, the mapping is not always one-to-one, and it is also not total. There are a couple of trace-units that do not have a corresponding part in the implementation, and there are also links to trace-units that are non-existent. Thus, as a first step in the preparation phase of this study, a small script was written for matching the references and units automatically, sorting out spelling errors and dangling links. Then, the result of the mapping has been verified and cross-checked by hand. This process yielded 70 units with a traceable transformation of Z code to $ADA$ code.

\medskip

The first part of the study deals with the question of relatedness between spe\-cifi\-cation-code pair measures. As we do not know whether the measures are normally distributed, three different statistical tests are used to assess the data: the Pearson's Correlation Coefficient,
the Spearman's Rank Correlation Coefficient, and Kendall's Tau Correlation Coefficient.
The Pearson's correlation coefficient ($R_P$) measures the degree of association between the variables,
assuming normal distribution of the values \cite[p.~212]{rees03}. Though this test might not
necessarily fail when the data is not normally distributed, the Pearson's test only looks for a linear
correlation. It might indicate no correlation even if the data is correlated in a non-linear
manner.
As the data might not be normally distributed, the Spearman's rank correlation coefficient
($R_S$) has been chosen \cite[p.~219]{rees03}. It is a non-parametric test of correlation and assesses how well
a monotonic function describes the association between the variables. This is done by ranking
the sample data separately for each variable.
Finally, the Kendall's robust correlation coefficient ($R_K$) is used as an alternative to the Spearman's
test \cite[p.~200]{fenton98}. It is also non-parametric and investigates the relationship
among pairs of data. However, it ranks the data relatively and is able to identify partial correlations.

\medskip

When a value of $|R|$ $\in$ $[0.8,~1.0]$ then it is interpreted to indicate a \emph{strong association}. When $|R|$ $\in$ $[0.5,~0.8)$ it is interpreted to indicate a \emph{moderate association}. When $|R|$ $\in$ $[0.0,~0.5)$ it is interpreted to indicate a \emph{weak association} (values rounded to the third decimal place).

\section{Correlation Tests}\label{sec:Correlation}

\begin{table}[t]
\begin{center}{\scriptsize
\begin{tabular}{|l||c|c|c|c|c|c|c|c|c|}\hline
\multicolumn{10}{|c|}{\textbf{Pearson Correlation $R_P$, $n=70$, $p \leq 0.033$)}}\\ \hline
\emph{Measure} & \emph{CL(M)} & \emph{CLC(M)} & \emph{CLCD(M)} & \emph{CLCE(M)} & \emph{CYC(M)} & \emph{KNOTS(M)} & \emph{FIN(M)} & \emph{FOUT(M)}  & \emph{SI(M)} \\ \hline \hline
$CC(\psi)$  & 0.806 & 0.681 & 0.343 & 0.797 & 0.749 & 0.842 & 0.258 & 0.866 & 0.255 \\ \hline
%$USE(\psi)$ & 0.798 & 0.688 & 0.371 & 0.790 & 0.752 & 0.821 & 0.278 & 0.842 & 0.279 \\ \hline
%$DEF(\psi)$ & 0.750 & 0.646 & 0.301 & 0.772 & 0.717 & 0.794 & 0.286 & 0.828 & 0.290 \\ \hline
$AND(\psi)$ & 0.538 & 0.507 & 0.369 & 0.519 & 0.586 & 0.477 & 0.366 & 0.482 & 0.384 \\ \hline
$OR(\psi)$  & 0.450 & 0.616 & 0.435 & 0.640 & 0.697 & 0.490 & 0.456 & 0.628 & 0.481 \\ \hline \hline
\multicolumn{10}{|c|}{\textbf{Spearman's Rank Correlation $R_S$, $n=70$, $p \leq 0.016$)}} \\ \hline
$CC(\psi)$  & 0.784 & 0.742 & 0.653 & 0.797 & 0.770 & 0.783 & 0.418 & 0.849 & 0.615 \\ \hline
%$USE(\psi)$ & 0.763 & 0.733 & 0.643 & 0.786 & 0.757 & 0.759 & 0.438 & 0.824 & 0.622 \\ \hline
%$DEF(\psi)$ & 0.742 & 0.698 & 0.595 & 0.775 & 0.735 & 0.745 & 0.402 & 0.802 & 0.610 \\ \hline
$AND(\psi)$ & 0.398 & 0.428 & 0.406 & 0.457 & 0.454 & 0.425 & 0.286 & 0.452 & 0.343 \\ \hline
$OR(\psi)$  & 0.697 & 0.731 & 0.699 & 0.760 & 0.748 & 0.704 & 0.497 & 0.774 & 0.619 \\ \hline
\multicolumn{10}{|c|}{\textbf{Kendall Robust Correlation $R_K$, $n=70$, $p \leq 0.025$)}}\\ \hline
$CC(\psi)$  & 0.586 & 0.544 & 0.462 & 0.595 & 0.577 & 0.623 & 0.300 & 0.686 & 0.448 \\ \hline
%$USE(\psi)$ & 0.569 & 0.544 & 0.466 & 0.594 & 0.573 & 0.593 & 0.311 & 0.657 & 0.453 \\ \hline
%$DEF(\psi)$ & 0.551 & 0.529 & 0.422 & 0.611 & 0.582 & 0.588 & 0.297 & 0.633 & 0.442 \\ \hline
$AND(\psi)$ & 0.289 & 0.308 & 0.286 & 0.343 & 0.334 & 0.341 & 0.200 & 0.356 & 0.241 \\ \hline
$OR(\psi)$  & 0.553 & 0.596 & 0.575 & 0.629 & 0.629 & 0.563 & 0.404 & 0.654 & 0.507 \\ \hline
\end{tabular}
}
\end{center}
\begin{center}{\scriptsize
\begin{tabular}{|l||c|c|c|c|c|c|c|c|c|}\hline
\multicolumn{10}{|c|}{\textbf{Pearson Correlation $R_P$, $n=70$, $p \leq 0.028$)}}\\ \hline
\emph{Measure} & \emph{CL(M)} & \emph{CLC(M)} & \emph{CLCD(M)} & \emph{CLCE(M)} & \emph{CYC(M)} & \emph{KNOTS(M)} & \emph{FIN(M)} & \emph{FOUT(M)}  & \emph{SI(M)} \\ \hline \hline
$v'_l(\psi)$ & 0.789 & 0.676 & 0.382 & 0.764 & 0.743 & 0.793 & 0.262 & 0.813 & 0.264 \\ \hline
$v'_u(\psi)$ & 0.787 & 0.661 & 0.358 & 0.758 & 0.739 & 0.794 & 0.259 & 0.808 & 0.262 \\ \hline
$DU(\psi)  $ & 0.799 & 0.642 & 0.285 & 0.777 & 0.736 & 0.817 & 0.279 & 0.833 & 0.282 \\ \hline
\multicolumn{10}{|c|}{\textbf{Spearman's Rank Correlation $R_S$, $n=70$, $p \leq 0.002$)}} \\ \hline
$v'_l(\psi)$ & 0.782 & 0.727 & 0.638 & 0.785 & 0.753 & 0.775 & 0.416 & 0.838 & 0.603 \\ \hline
$v'_u(\psi)$ & 0.764 & 0.695 & 0.602 & 0.762 & 0.725 & 0.785 & 0.363 & 0.824 & 0.556 \\ \hline
$DU(\psi)  $ & 0.767 & 0.682 & 0.593 & 0.777 & 0.728 & 0.784 & 0.377 & 0.832 & 0.600 \\ \hline
\multicolumn{10}{|c|}{\textbf{Kendall Robust Correlation $R_K$, $n=70$, $p \leq 0.003$)}}\\ \hline
$v'_l(\psi)$ & 0.603 & 0.543 & 0.460 & 0.602 & 0.583 & 0.628 & 0.305 & 0.691 & 0.441 \\ \hline
$v'_u(\psi)$ & 0.565 & 0.502 & 0.431 & 0.552 & 0.533 & 0.619 & 0.262 & 0.655 & 0.402 \\ \hline
$DU(\psi)  $ & 0.567 & 0.518 & 0.431 & 0.584 & 0.558 & 0.613 & 0.280 & 0.659 & 0.430 \\ \hline
\end{tabular}
}
\end{center}
\caption{Pearson's, Spearman's and Kendall's correlation for size- and structure based Z measures.
}
\label{tab:Corr-SizeBased}
\end{table}

After data preparation, the study looked for linear or at least partial
correlations between the sets of measures. At first, classical size-based measures are considered, and
Table \ref{tab:Corr-SizeBased} (upper part) summarizes the results. The p-values for testing the hypothesis of no correlation against the alternative that there is a nonzero correlation are less than $0.05$ for all tests. The table also shows that there is a moderate to strong relation between $CC(\psi)$ and the measures of $CL(M)$, $KNOTS(M)$ and $FOUT(M)$. The correlation values of the tests are quite similar, but there are a couple of exceptions. Compared to the Pearson test, the Spearman's rank test shows a higher correlation between most of the size-based measures and the Count Line Declarative $CLCD(M)$ measure, indicating that there might be a non-linear correlation between them. However, the Kendall's test shows weak correlation for most of the measures again. A similar situation can be observed for the measure of $FIN(M)$. Here, the Spearman test shows a slightly higher correlation than the other two tests, but it still falls into the weak association class. Interesting are the differences between the tests for the $SI(M)$ measure.
The correlation to the size-based measures is not strong, but $SI(M)$ is calculated by also using the square of $FOUT(M)$, and this non-linear tendency can be seen in the slightly higher values of the Spearman tests. And yet another issue can
be observed: cyclomatic complexity is (although only moderately) influenced by the number of logical OR connections
in the specification. As cyclomatic complexity is related to the number of paths through the program, this observation seems also to be consistent to the use of or-combined predicates in a Z specification.

%\begin{figure}[t]
%  \centerline{\psfig{figure=SemanticsBased.eps,width=0.9\textwidth}}
%  \caption{Scatter plots for the correlation between logical complexity $v'_l(\psi)$ and $FOUT(M)$ %(left side) and for the correlation between coverage $Cov(\psi)$ and Count Line $CL(M)$ ($n=70$).
%    \label{fig:Correlation}}
%\end{figure}

\medskip
\begin{table}[thb]
\begin{center}{\scriptsize
\begin{tabular}{|l|l||c|c|c||c|c|c||c|c|c|}\hline
\multicolumn{11}{|c|}{\textbf{Semantics-based Correlation, $n=70$}}\\ \hline
\multicolumn{2}{|c||}{} & \multicolumn{3}{|c||}{Pearson}  & \multicolumn{3}{|c||}{Spearman} & \multicolumn{3}{|c|}{Kendall} \\ \cline{3-11}
\multicolumn{2}{|c||}{} & \emph{$Cov(\psi)$} &\emph{$O(\psi)$} & \emph{$\chi(\psi)$} & \emph{$Cov(\psi)$} &\emph{$O(\psi)$} & \emph{$\chi(\psi)$} & \emph{$Cov(\psi)$} &\emph{$O(\psi)$} & \emph{$\chi(\psi)$} \\ \hline \hline
CL(M)    & R  & 0.104       & -.623       & 0.646       & 0.054       & -.616       & 0.686 & 0.042       & -.495       & 0.480 \\ \cline{2-11}
         & p  & {\bf 0.391} & 0.000       & 0.000       & {\bf 0.660} & 0.000       & 0.000 & {\bf 0.624} & 0.000       & 0.000 \\ \hline
CLC(M)   & R  & 0.184       & -.466       & 0.414       & 0.186       & -.480       & 0.541 & 0.146       & -.364       & 0.379 \\ \cline{2-11}
         & p  & {\bf 0.127} & 0.000       & 0.000       & {\bf 0.124} & 0.000       & 0.000 & {\bf 0.085} & 0.000       & 0.000 \\ \hline
CLCD(M)  & R  & 0.229       & -.215       & 0.116       & 0.243       & -.369       & 0.433 & 0.190       & -.280       & 0.310 \\ \cline{2-11}
         & p  & {\bf 0.057} & {\bf 0.075} & {\bf 0.340} & 0.042       & 0.002       & 0.000 & 0.026       & 0.003       & 0.000 \\ \hline
CLCE(M)  & R  & 0.126       & -.559       & 0.546       & 0.110       & -.587       & 0.636 & 0.079       & -.461       & 0.440 \\ \cline{2-11}
         & p  & {\bf 0.300} & 0.000       & 0.000       & {\bf 0.363} & 0.000       & 0.000 & {\bf 0.355} & 0.000       & 0.000 \\ \hline
CYC(M)   & R  & 0.148       & -.534       & 0.509       & 0.137       & -.531       & 0.590 & 0.103       & -.416       & 0.412 \\ \cline{2-11}
         & p  & {\bf 0.221} & 0.000       & 0.000       & {\bf 0.258} & 0.000       & 0.000 & {\bf 0.234} & 0.000       & 0.000 \\ \hline
KNOTS(M) & R  & 0.062       & -.587       & 0.637       & -.018       & -.629       & 0.721 & -.034       & -.530       & 0.542 \\ \cline{2-11}
         & p  & {\bf 0.613} & 0.000       & 0.000       & {\bf 0.884} & 0.000       & 0.000 & {\bf 0.708} & 0.000       & 0.000 \\ \hline
FIN(M)   & R  & 0.150       & -.224       & 0.212       & 0.245       & -.170       & 0.267 & 0.184       & -.132       & 0.193 \\ \cline{2-11}
         & p  & {\bf 0.216} & {\bf 0.062} & {\bf 0.078} & 0.041       & {\bf 0.160} & 0.026 & 0.034       & {\bf 0.164} & 0.026 \\ \hline
FOUT(M)  & R  & 0.096       & -.572       & 0.582       & 0.046       & -.599       & 0.722 & -.008       & -.479       & 0.541 \\ \cline{2-11}
         & p  & {\bf 0.430} & 0.000       & 0.000       & {\bf 0.704} & 0.000       & 0.000 & {\bf 0.930} & 0.000       & 0.000 \\ \hline
SI(M)    & R  & 0.123       & -.235       & 0.227       & 0.220       & -.406       & 0.480 & 0.159       & -.315       & 0.339 \\ \cline{2-11}
         & p  & {\bf 0.309} & 0.050       & {\bf 0.059} & {\bf 0.067} & 0.000       & 0.000 & {\bf 0.063} & 0.001       & 0.000 \\ \hline
\end{tabular}
}
\end{center}
\caption{Pearson, Spearman and Kendall for semantics-based measures.}
\label{tab:Corr-Semantics}
\end{table}

\begin{table}[t]
\begin{center}{\scriptsize
\begin{tabular}{|c||c|c|c|c|c|c|}\hline
\multicolumn{7}{|c|}{\textbf{Results of the backward elimination procedure (values $P \leq 0.4$)}}\\ \hline\hline
\multicolumn{2}{|c|}{\emph{Paramter}}    & \emph{CL(M)} & \emph{CLE(M)} & \emph{CYC(M)} & \emph{KNOTS(M)} & \emph{FOUT(M)} \\ \hline
\multicolumn{2}{|c|}{Adjusted R-Square}  & 0.720   & 0.680   & 0.620   & 0.760   & 0.840     \\ \hline
\multicolumn{2}{|c|}{Significance F}     & 5E-18   & 2E-16   & 5E-14   & 7E-20   & 1.5E-25 \\ \hline\hline
          & $CC(\psi)$                   & 0.001   & 4E-4    & 0.003   & 1E-6    & 3E-11     \\ \cline{2-7}
          & $v'_l(\psi)$                 & ------  & -----   & -----   & -----   & 0.270     \\ \cline{2-7}
          & $v'_u(\psi)$                 & ------  & 0.005   & -----   & 0.004   & 0.000     \\ \cline{2-7}
          & $DU(\psi)$                   & ------  & -----   & -----   & -----   & -----     \\ \cline{2-7}
          & $O(\psi)$                    & ------  & -----   & -----   & -----   & -----     \\ \cline{2-7}
$P-Value$ & $Cov(\psi)$                  & ------  & -----   & 0.280   & -----   & -----     \\ \cline{2-7}
          & $\chi(\psi)$                 & ------  & -----   & -----   & -----   & -----     \\ \cline{2-7}
          & $AND(\psi)$                  & 7E-5    & -----   & -----   & 0.030   & -----     \\ \cline{2-7}
          & $OR(\psi)$                   & 5E-4    & 0.001   & 4E-4    & -----   & 6E-5      \\ \cline{2-7}
          & $DEF(\psi)$                  & ------  & 0.070   & -----   & 0.020   & 1E-5      \\ \cline{2-7}
          & $USE(\psi)$                  & 0.034   & -----   & -----   & 0.320   & -----     \\ \hline
\end{tabular}
}
\end{center}
\caption{Adjusted R-Square, Significance F and P-Values after applying the backward elimination procedure for maximum model identification. P-Values higher than $0.4$ are represented by dashes.}
\label{tab:Regression}
\end{table}

In a second step, structure-based measures have been looked at. Table \ref{tab:Corr-SizeBased} (lower part) summarizes the results. Again, the p-values are less than $0.05$ for all tests. The correlations are not as strong as with the pure size-based measures -- with one exception: the structure-based measures seem to strongly influence the $FOUT(M)$ count.
%Figure \ref{fig:Correlation} (left side) shows a scatter plot for the correlation between logical complexity $v'_l(\psi)$ and $FOUT(M)$ and a trend is recognized. The higher the number of dependencies, the higher is the number of control and data flows to other $ADA$ modules.
The other structure-based measures moderately to strongly influence the complexity measures $CYC(M)$ and $KNOTS(M)$.
This seems to be inherent, as these measures are counting dependencies within and between the schemas. The correlation to the other $ADA$-related measures in also moderate to strong. Only the measures of $CLCD(M)$, $FIN(M)$ and $SI(M)$ do have weak correlations.

\medskip

In the case of semantics-based measures the picture has to be looked at in a more differentiated way (see Table \ref{tab:Corr-Semantics}). At first, most of the results of the tests concerning \emph{Coverage} are statistically not significant (higher $p$ values are shown in bold numerals). The tests indicate no correlation between the $ADA$-based measures and \emph{Coverage}, but the chance is high that this is wrong. In this situation scatter plots have been used to gain a better understanding of the results, but the plots
%, like those in Figure \ref{fig:Correlation} (right side)
confirmed the results of no correlation at all. The other tests indicate weak to moderate relations for \emph{Overlap} and \emph{Coupling}, but another point is interesting. \emph{Overlap} and \emph{Coupling} have different leading signs. This might partially be explained by the fact that overlap is an indicator for crispness. It is high when the schema is not strongly related to other parts of the specification. And coupling is higher when there are more relations to other specification schemas. An increase in the value of one measure leads to a decrease of the other measure.

\medskip

To summarize, there is only weak to moderate relation between the Z-based measures and $CLCD(M)$, $FIN(M)$, and $SI(M)$. But, though not exclusively, there is some moderate to strong correlation between the Z-based and the other imple\-mentation-based measures. When just focusing, for example, on $CL(M)$, $CYC(M)$, $KNOTS(M)$, and $FOUT(M)$ and taking moderate to strong correlations into account, then the following can be observed: Firstly, they are all influenced by structure-based measures. Secondly, especially $CL(M)$, $KNOT(M)$ and $FOUT(M)$ do have strong correlations to the Z measures. The next section now uses the Z measures to provide regression formulas for the most suitable $ADA$-based measures.

\section{Prediction Models}\label{sec:Prediction}

According to a rule of thumb in regression \cite[p.3]{lars08}, the appropriate number of independent variables for a prediction is not more than one fifth of the sample size. Thus, the eleven Z measures presented in Section \ref{sec:Measures} can be considered to be sufficient and they are all selected to form the maximum model for 70 observations in this study. Among several systematic methods for restricting the maximum model, a backward elimination procedure  \cite[p.8]{lars08} with a threshold of $0.4$ for the P-values is selected. This means that a maximum regression model with all eleven independent variables is built. Then all the variables with a P-value of more than $0.4$ are eliminated. Then, again, another regression model with the reduced number of variables is built, iterating until there is no variable with a P-value higher than $0.4$.

\medskip

Table \ref{tab:Regression} summarizes the final result of this procedure for the five remaining code metrics (as measures with a P-Value higher than $0.4$ have been eliminated). The table, for example, shows that for the calculation of the cyclomatic complexity $CYC(M)$ of an $ADA$ module, $CC(\psi)$, $Cov(\psi)$ and $OR(\psi)$ are best for being used in the regression formula.
The level of confidence can be explained by the values of \emph{Significance-F}. If the level of acceptable confidence should be 95\% and higher, then all the code metrics with F-values of less than $0.05$ can be considered predictable using the metrics in Z. All the values for $F$ in Table \ref{tab:Regression} show that there is a high reliability on the results of the regressions.
The value of \emph{Adjusted R-Square} can be interpreted as an indicator for the precision level of the prediction.
In our case the values are between $0.620$ and $0.840$, indicating that the regression models are relatively precise for $FOUT(M)$, $KNOTS(M)$ and $CL(M)$ and even more precise for $CLE(M)$ and $CYC(M)$.
With these values at hand it makes sense to predict code metrics, and the resulting formulas are as follows:

\begin{center}{\small
$CL(M)  = 3.099  CC(\psi) - 1.237  USE(\psi) + 2.557  AND(\psi) - 41.735  OR(\psi) - 9.873$
$CLCE(M)  =  0.516 CC(\psi) - 0.003  v'_u(\psi) - 0.477  DEF(\psi) + 5.458  OR(\psi) + 5.819$
$CYC(M) = 0.015 CC(\psi) + 4.349 Cov(\psi) - 2.107 O(\psi) + 1.082 OR(\psi) + 1.666$
$KNOTS(M) = 0.121  CC(\psi) - 0.001  v'_u(\psi) - 0.017  USE(\psi) - 0.092  DEF(\psi) + 0.027  AND(\psi) - 0.882$
$FOUT(M) = 0.198  CC(\psi) - 0.107  v'_l(\psi) - 0.001  v'_u(\psi) - 0.211  DEF(\psi) +  1.220  OR(\psi) + 0.344$
}\end{center}

%\begin{figure}[t]
%  \centerline{\psfig{figure=Regression.eps,width=0.8\textwidth}}
%  \caption{Comparison between the real and the predicted measures (according to the regression %formulas) concerning the number of lines of code $CL(M)$ and $FOUT(M)$ ($n= 70)$.
%    \label{fig:Regression}}
%\end{figure}

%Figure \ref{fig:Regression} depicts the precision of the prediction models for the measures of $CL(M)$ and $FOUT(M)$. For all 70 samples it compares the real and the predicted values of the measures and visualizes the (not always, but mostly) usually small gaps between them. $FOUT(M)$ had the highest precision level, and thus it is not surprising that it has less outliers than the $CL(M)$ measures. But also in the case of $CL(M)$ there are only a couple of values which are far outside the expected range.

\section{Threats to Validity}\label{sec:Validity}

With the results of the study the question of validity arises.
Considering internal validity, single group and multiple group threats, as well as social threats
cannot arise. The only threat that might have an impact on the outcome of the study is the software used to generate and calculate the measures. The software components involved are the $CZT$ parser \cite{malik11},
the slicing environment $ViZ$ \cite{boll07a}, $Matlab~R2007b$, $Microsoft~Excel~2010$ and $SPSS$ $14$:$0$. $Excel$ is a standard spreadsheet application. $Matlab$ and $SPSS$ are numerical computer environments used for the statistical analysis. Both tools have been used alternately
to verify the results of the analysis. It is very unlikely that the data from both environments is
erroneous. The $CZT$ parser is being developed as a $SourceForge$ project since 2003 and it is available in a stable release.  The slicing environment $ViZ$ has been developed in the year 2003 and it is also part of a couple of extensions which led to systematic validations during development.

Concerning the selection validity, the publicly available schemas and $ADA$ procedures and functions have been chosen with care, following the links provided by the developers. It is important to note that the specification used in this study had to be modified a bit in order to be accepted by the $CZT$ parser. This meant to introduce some hard spaces and, eventually, also to replace the ``$\sdef$'' symbol by the ``$==$'' sign. In order to rule out the possibility of coincidental changes of line breaks or
identifier names, both files were again compared afterwards, using a professional file-compare software.

\section{Conclusion}\label{sec:Conclusion}

In this study, consisting of 70 experimental subjects, the feasibility of confidently
predicting software measures based on formal specifications has been demonstrated. The correlations found between the different size-, structure-, and semantics-based measures and the implementation metrics promise of being able to predict size and complexity attributes as well as enables to estimate likely costs and efforts.

The study describes only the first link in the chain of associations between the documents created during software development, but it confirms the observations of Samson et.al.~\cite{sams87} who conducted a similar study (with 9 experimental subjects) several years ago. Specification-based measures are not difficult to calculate, thus they can and also \emph{should} be collected at the beginning of a project. The results of the study indicate that it pays off.

\bibliographystyle{eptcs}
\bibliography{fmds12}

\begin{thebibliography}{10}
\providecommand{\bibitemdeclare}[2]{}
\providecommand{\surnamestart}{}
\providecommand{\surnameend}{}
\providecommand{\urlprefix}{Available at }
\providecommand{\url}[1]{\texttt{#1}}
\providecommand{\href}[2]{\texttt{#2}}
\providecommand{\urlalt}[2]{\href{#1}{#2}}
\providecommand{\doi}[1]{doi:\urlalt{http://dx.doi.org/#1}{#1}}
\providecommand{\bibinfo}[2]{#2}

\bibitemdeclare{phdthesis}{boll04}
\bibitem{boll04}
\bibinfo{author}{Andreas \surnamestart Bollin\surnameend}
  (\bibinfo{year}{2004}): \emph{\bibinfo{title}{Specification Comprehension --
  Reducing the Complexity of Specifications}}.
\newblock Ph.D. thesis, \bibinfo{school}{University of Klagenfurt}.

\bibitemdeclare{article}{boll07a}
\bibitem{boll07a}
\bibinfo{author}{Andreas \surnamestart Bollin\surnameend}
  (\bibinfo{year}{2007}): \emph{\bibinfo{title}{Concept Location in Formal
  Specifications}}.
\newblock {\sl \bibinfo{journal}{Journal of Software Maintenance and Evolution:
  Research and Practice}} \bibinfo{volume}{Manuscript submitted Jan. 2007},
  \doi{10.1002/smr.363}.

\bibitemdeclare{inproceedings}{Bollin10}
\bibitem{Bollin10}
\bibinfo{author}{Andreas \surnamestart Bollin\surnameend}
  (\bibinfo{year}{2010}): \emph{\bibinfo{title}{{Slice-based Formal
  Specifiation Measures -- Mapping Coupling and Cohesion Measures to Formal
  Z}}}.
\newblock In \bibinfo{editor}{C{\`e}sar \surnamestart Mu{\~n}oz\surnameend},
  editor: {\sl \bibinfo{booktitle}{Proceedings of the Second NASA Formal
  Methods Symposium}}, \bibinfo{series}{NASA/CP-2010-216215},
  \bibinfo{publisher}{NASA, Langley Research Center}, pp.
  \bibinfo{pages}{24--34}.

\bibitemdeclare{techreport}{chang94}
\bibitem{chang94}
\bibinfo{author}{Juei \surnamestart Chang\surnameend} \&
  \bibinfo{author}{Debra~J. \surnamestart Richardson\surnameend}
  (\bibinfo{year}{1994}): \emph{\bibinfo{title}{{Static and Dynamic
  Specification Slicing}}}.
\newblock \bibinfo{type}{Technical Report}, \bibinfo{institution}{Department of
  Information and Computer Science, University of California}.

\bibitemdeclare{manual}{chap09}
\bibitem{chap09}
\bibinfo{author}{Rod \surnamestart Chapman\surnameend} (\bibinfo{year}{2009}):
  \emph{\bibinfo{title}{The Tokeneer ID Station -- Overview and Readers Guide.
  S.P1229.81.8. Issue: 1.4}}.
\newblock \bibinfo{organization}{Praxis High Integrity Systems}.

\bibitemdeclare{book}{fenton98}
\bibitem{fenton98}
\bibinfo{author}{Norman~E. \surnamestart Fenton\surnameend} \&
  \bibinfo{author}{Shari~Lawrence \surnamestart Pfleeger\surnameend}
  (\bibinfo{year}{1989}): \emph{\bibinfo{title}{{Software Metrics}}},
  \bibinfo{edition}{2nd} edition.
\newblock \bibinfo{publisher}{Thompson Press}.

\bibitemdeclare{inproceedings}{har97}
\bibitem{har97}
\bibinfo{author}{Mark \surnamestart Harman\surnameend},
  \bibinfo{author}{Margaret \surnamestart Okulawon\surnameend},
  \bibinfo{author}{Bala \surnamestart Sivagurunathan\surnameend} \&
  \bibinfo{author}{Sebastian \surnamestart Danicic\surnameend}
  (\bibinfo{year}{1997}): \emph{\bibinfo{title}{Slice-based measurement of
  coupling}}.
\newblock In: {\sl \bibinfo{booktitle}{Proceedings of the ICSE workshop on
  Process Modelling and Empirical Studies of Software Evolution. Boston,
  Massachusetts}}, \bibinfo{publisher}{IEEE Computer Society},
  \bibinfo{address}{Los Alamitos, CA, USA}, pp. \bibinfo{pages}{28--32}.

\bibitemdeclare{article}{hen81}
\bibitem{hen81}
\bibinfo{author}{Sally~M. \surnamestart Henry\surnameend} \&
  \bibinfo{author}{Dennis~G. \surnamestart Kafura\surnameend}
  (\bibinfo{year}{1981}): \emph{\bibinfo{title}{Software structure metrics
  based on information flow}}.
\newblock {\sl \bibinfo{journal}{IEEE Transactions on Software Engineering}}
  \bibinfo{volume}{7}(\bibinfo{number}{5}), pp. \bibinfo{pages}{510--518},
  \doi{10.1109/TSE.1981.231113}.

\bibitemdeclare{article}{Jorgensen04}
\bibitem{Jorgensen04}
\bibinfo{author}{Magne \surnamestart J{\o}rgensen\surnameend}
  (\bibinfo{year}{2004}): \emph{\bibinfo{title}{{A review of studies on expert
  estimation of software development effort}}}.
\newblock {\sl \bibinfo{journal}{Journal of Systems and Software}}
  \bibinfo{volume}{70}(\bibinfo{number}{1--2}), pp. \bibinfo{pages}{37--60},
  \doi{10.1016/S0164-1212(02)00156-5}.

\bibitemdeclare{manual}{lars08}
\bibitem{lars08}
\bibinfo{author}{Pia~Veldt \surnamestart Larsen\surnameend}
  (\bibinfo{year}{2008}): \emph{\bibinfo{title}{ST111: Regression and analysis
  of variance -- Module 8: Selecting regression models}}.
\newblock \bibinfo{organization}{Manual from
  statmaster.sdu.dk/courses/st111/module08, Syddansk University, Department of
  Statistics}.

\bibitemdeclare{article}{malik11}
\bibitem{malik11}
\bibinfo{author}{Petra \surnamestart Malik\surnameend} (\bibinfo{year}{2011}):
  \emph{\bibinfo{title}{A retrospective on CZT}}.
\newblock {\sl \bibinfo{journal}{{Software -- Practice and Experience}}}
  \bibinfo{volume}{41}(\bibinfo{number}{2}), pp. \bibinfo{pages}{179--188},
  \doi{10.1002/spe.1015}.

\bibitemdeclare{inproceedings}{mitt03}
\bibitem{mitt03}
\bibinfo{author}{Roland~T. \surnamestart Mittermeir\surnameend} \&
  \bibinfo{author}{Andreas \surnamestart Bollin\surnameend}
  (\bibinfo{year}{2003}): \emph{\bibinfo{title}{Demand-driven Specification
  Partitioning}}.
\newblock In: {\sl \bibinfo{booktitle}{Proceedings of the 5th Joint Modular
  Languages Conference, JMLC'03}}, pp. \bibinfo{pages}{241--253},
  \doi{10.1007/978-3-540-45213-3\_30}.

\bibitemdeclare{inproceedings}{oda93}
\bibitem{oda93}
\bibinfo{author}{Tomohiro \surnamestart Oda\surnameend} \&
  \bibinfo{author}{Keijiri \surnamestart Araki\surnameend}
  (\bibinfo{year}{1993}): \emph{\bibinfo{title}{Specification slicing in a
  formal methods software development}}.
\newblock In: {\sl \bibinfo{booktitle}{Seventeenth Annual International
  Computer Software and Applications Conference}}, \bibinfo{series}{IEEE
  Computer Socienty Press}, pp. \bibinfo{pages}{313--319},
  \doi{10.1109/CMPSAC.1993.404234}.

\bibitemdeclare{inproceedings}{ott89}
\bibitem{ott89}
\bibinfo{author}{Linda~M. \surnamestart Ott\surnameend} \&
  \bibinfo{author}{Jeffrey~J. \surnamestart Thus\surnameend}
  (\bibinfo{year}{1989}): \emph{\bibinfo{title}{{The Relationship between
  Slices and Module Cohesion}}}.
\newblock In: {\sl \bibinfo{booktitle}{11th International Conference on
  Software Engineering}}, \bibinfo{publisher}{IEEE Computer Society},
  \bibinfo{address}{Los Alamitos, CA, USA}, pp. \bibinfo{pages}{198--204},
  \doi{10.1145/74587.74614}.

\bibitemdeclare{book}{putnam03}
\bibitem{putnam03}
\bibinfo{author}{Lawrence~H. \surnamestart Putnam\surnameend} \&
  \bibinfo{author}{Ware \surnamestart Myers\surnameend} (\bibinfo{year}{2003}):
  \emph{\bibinfo{title}{{Five Core Metrics: The Intelligence Behind Successful
  Software Management}}}.
\newblock \bibinfo{publisher}{Dorset House}.

\bibitemdeclare{book}{rees03}
\bibitem{rees03}
\bibinfo{author}{D.~G. \surnamestart Rees\surnameend} (\bibinfo{year}{2003}):
  \emph{\bibinfo{title}{{Essential Statistics}}}, \bibinfo{edition}{4th}
  edition.
\newblock \bibinfo{publisher}{Chapman \& Hall}.

\bibitemdeclare{inproceedings}{sams87}
\bibitem{sams87}
\bibinfo{author}{W.B. \surnamestart Samson\surnameend},
  \bibinfo{author}{Denis~G. \surnamestart Nevill\surnameend} \&
  \bibinfo{author}{P.I. \surnamestart Dugard\surnameend}
  (\bibinfo{year}{1987}): \emph{\bibinfo{title}{Predictive software metrics
  based on a formal specification}}.
\newblock In: {\sl \bibinfo{booktitle}{Information and Software Technology}},
  {\sl \bibinfo{series}{5}}~\bibinfo{volume}{29}, pp.
  \bibinfo{pages}{242--248}.

\bibitemdeclare{inproceedings}{shep90}
\bibitem{shep90}
\bibinfo{author}{Martin~J. \surnamestart Shepperd\surnameend} \&
  \bibinfo{author}{Darrel~C. \surnamestart Ince\surnameend}
  (\bibinfo{year}{1990}): \emph{\bibinfo{title}{The use of metrics in the early
  detection of design errors}}.
\newblock In: {\sl \bibinfo{booktitle}{Proceedings of the European Software
  Engineering Conference '90}}, pp. \bibinfo{pages}{67--85}.

\bibitemdeclare{book}{sneed10}
\bibitem{sneed10}
\bibinfo{author}{Harry~M. \surnamestart Sneed\surnameend},
  \bibinfo{author}{Richard \surnamestart Seidl\surnameend} \&
  \bibinfo{author}{Manfred \surnamestart Baumgartner\surnameend}
  (\bibinfo{year}{2010}): \emph{\bibinfo{title}{{Software in Zahlen}}}.
\newblock \bibinfo{publisher}{Carl Hanser Verlag}.

\bibitemdeclare{mastersthesis}{tab11}
\bibitem{tab11}
\bibinfo{author}{Abdollah \surnamestart Tabareh\surnameend}
  (\bibinfo{year}{2011}): \emph{\bibinfo{title}{{Predictive Software Measures
  Based on Formal Z Specifications}}}.
\newblock Master's thesis, \bibinfo{school}{University of Gothenburg -
  Department of Computer Science and Engineering}.

\bibitemdeclare{inproceedings}{weiser82}
\bibitem{weiser82}
\bibinfo{author}{Mark \surnamestart Weiser\surnameend} (\bibinfo{year}{1982}):
  \emph{\bibinfo{title}{{Program slicing}}}.
\newblock In: {\sl \bibinfo{booktitle}{Proceedings of the $5^{th}$
  International Conference on Software Engineering}}, \bibinfo{publisher}{IEEE
  Press}, \bibinfo{address}{Piscataway, NJ, USA}, pp.
  \bibinfo{pages}{439--449}.

\end{thebibliography}

\end{document}